# A Computational Model for the Capillary Flow between Parallel Plates

Mohammad Naghashnejad, Hamidreza Shabgard*

*School of Aerospace and Mechanical Engineering, University of Oklahoma, Norman, OK, USA*

*Corresponding Author Phone: +1-405-3255753, E-mail: shabgard@ou.edu*

**Abstract**

A computational fluid dynamics (CFD) model is developed to simulate the dynamics of meniscus formation and capillary flow between vertical parallel plates. The arbitrary Lagrangian-Eulerian (ALE) approach was employed to predict and reconstruct the exact shape of the meniscus. The model was used to simulate the rise of water and the evolution of the meniscus in vertical channels with various spacing values of 0.5 mm, 0.7 mm, and 1 mm. The validity of the model was established by comparing the steady-state capillary height and the meniscus profile with analytical solutions. The developed model presents a novel approach for simulation of capillary flow accounting for the detailed hydrodynamic phenomena that cannot be captured by analytical models.

## 1. Introduction

Liquid flows driven by capillary action appear in a wide range of applications in natural and industrial systems, including water transport in plants, heat pipes and spacecraft Propellant Management Devices (PMDs) [1]. Capillary action is the spontaneous movement of liquid within the voids formed by solid surfaces due to the interaction of adhesion, cohesion, and surface tension forces. The adhesion between the liquid and solid walls along with the surface tension force (cohesion between the liquid molecules) at the liquid-gas interface create the driving force for penetration of the liquid column inside the capillary conduit. This force is countered by the gravity and viscous forces as the liquid flows through the capillary conduit. The counteraction of the gravity and the adhesion-cohesion forces at the interface creates the meniscus shape shown in Fig. 1. The surface tension force acting along the meniscus induces a pressure jump across the interface that can be determined using the Young-Laplace equation.

Flow in capillary conduits has been extensively studied. The earliest investigations of the transient rise of liquids in a capillary channel were performed by Lucas [2] and Washburn [3]. In their analyses, a theoretical equation known as the Lucas-Washburn equation (LWE) was developed by balancing surface tension, viscous and gravity forces acting on the fluid control volume. The LWE does not include the inertial effects in the momentum balance. As such, it fails to predict the early stages of liquid penetration, where the gravitational and viscous forces are relatively small, and the inertial effects are the primary factor to balance the capillary force. Rideal [4] and Bosanquet [5] considered the contribution of the inertial effects on the liquid rise in capillary channels. The inclusion of inertia eliminates a non-physical initial velocity observed in the early theoretical models. Levin et al. [6] assumed a more accurate pressure field in the reservoir near the entry region of the circular capillary and developed one of the most comprehensive analytical models available in the literature. Several other researchers have developed similar analyses for non-circular capillaries and parallel plates with finite length and considered other effective parameters such as entrance pressure difference and dynamic contact angle [7-9].

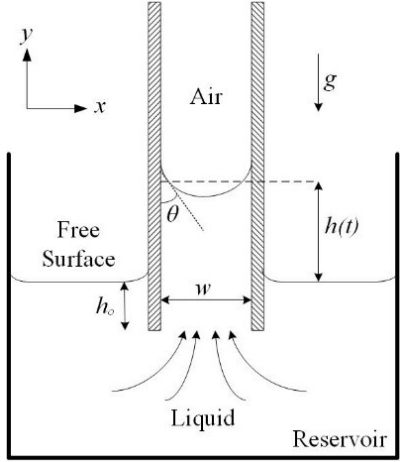

Fig. 1. Schematic of the liquid-rise in a vertical capillary.

Compared to analytical models, only a few computational fluid dynamics (CFD) studies have been carried out to study the capillary flow. Among available CFD models, the volume of fluid (VOF) has been one of the most popular methods [10-11]. In this method, the liquid-gas interface is marked by the cells that have a volume fraction between 0 and 1, and the exact interface information is discarded because of discrete volume fraction data. More recently, Grunding et al. [12] conducted a computational analysis of the problem of a liquid rising between parallel plates using various numerical approaches including arbitrary Lagrangian-Eulerian (ALE) method. They assumed that the meniscus was a circular arc and remained unchanged during the capillary penetration.

Review of the literature reveals that despite the extensive amount of research conducted on capillary flow, there is a lack of models that can predict the transient evolution of the meniscus profile and directly track the sharp liquid-gas interface. The objective of this study is to develop a CFD model for simulation of the meniscus formation dynamics and capillary flow behavior of liquid column between two vertical parallel plates. The two-dimensional transient conservation equations for mass and momentum were solved with the ALE method in a finite volume formulation using commercial CFD package ANSYS Fluent. A dynamic mesh method was implemented to directly track the interface with no need for implicit interface tracking schemes.

## 2. Problem Description

A schematic of the physical problem is shown in Fig. 1. The capillary channel was formed by partially submerging two vertical parallel aluminum plates with a wall spacing of $w$ in a liquid water reservoir. The submerged section of the capillary channel was 3 mm long and the wall spacing was varied from 0.5 mm to 1 mm. The reservoir length in the x and y directions was 22.7 mm and 10 mm respectively. The initial height of the liquid in the reservoir was 8 mm and the remaining space was filled with air at atmospheric pressure. After the liquid comes in contact with the channel walls, a contact angle $\theta$ is established between the two. The adhesion between the liquid and solid walls together with the cohesion forces between the water molecules at the liquid-air interface lead to formation of a meniscus at the liquid-air interface. The liquid starts to rise through the capillary due to the interaction between adhesion, inertia, gravity, and viscous forces. The rise of the liquid in the capillary channel induces the liquid level in the reservoir to recede. The variations of the liquid level in the reservoir are small compared to that in the capillary channel due to the significantly greater free surface in the reservoir. Both air and water were assumed to be incompressible and Newtonian fluids with constant properties shown in Table 1.

Table 1. Physical properties of water and air.

| Fluid | Water | Air |
|---|---|---|
| Density $\rho$ (kg/m3) | 998.2 | 1.225 |
| Dynamic viscosity $\mu$ (kg/m.s) | 1.003e-03 | 1.7894e-05 |
| Surface tension $\sigma$ (N/m) | 0.0728 | – |

## 3. Computational Model

Transient conservation equations for mass and momentum, respectively, can be written as:

$$\frac{\partial \rho}{\partial t} + \nabla \cdot (\rho \vec{v}) = 0 \tag{1}$$

$$\frac{\partial}{\partial t}(\rho \vec{v}) + \nabla \cdot (\rho \vec{v} \vec{v}) = -\nabla p + \nabla \cdot [\mu(\nabla \vec{v} + \nabla \vec{v}^T)] + \rho \vec{g} \tag{2}$$

where $t$ is time, $\rho$, $\mu$, and $\vec{v}$ are the density, dynamic viscosity, and velocity vector, respectively, $p$ is pressure, and $\vec{g}$ is gravity. The VOF model was employed to capture the liquid-air interface in the reservoir. The equation governing the evolution of the water volume fraction distributions in the reservoir is:

$$\frac{\partial \alpha_l}{\partial t} + \vec{v} \cdot \nabla \alpha_l = 0 \tag{3}$$

where $\alpha_l$ is the volume fraction of water. The volume fraction of the air is determined from $\alpha_g = 1 - \alpha_l$.

A dynamic mesh method was implemented to directly track the interface with no need for implicit interface tracking schemes. This technique is useful in simulation of the fluid flow problems with moving boundaries. For an arbitrary control volume $V$ with a moving boundary of $\partial V$, the integral form of the conservation equation for a generic scalar $\lambda$ can be expressed as [14]:

$$\frac{d}{dt}\int_V \rho \lambda dV + \int_{\partial V} \rho \lambda (\vec{v} - \vec{v}_m) \cdot d\vec{A} = \int_{\partial V} \Psi \nabla \lambda \cdot d\vec{A} + \int_V S_\lambda dV \tag{4}$$

where $\vec{v}_m$ is the moving velocity of the mesh, $\Psi$ is the diffusion coefficient and $S_\lambda$ is the source term.

The above governing equations were solved under the assumption of laminar flow and the SIMPLE algorithm was used for velocity-pressure coupling. The pressure, velocity, and volume fraction at the cell interfaces were interpolated using the PRESTO!, power-law and piecewise-linear interpolation schemes, respectively. At each time step, the capillary pressure differential resulting from the curvature of the water-air interface was determined using the Laplace's equation. The pressure profile was then applied to the interfacial faces and the governing equations were solved. The calculated velocity at the interfacial cells were used to update the position of the interfacial nodes. A new pressure profile was determined based on the updated meniscus shape and was applied as the boundary condition for the new time step. Due to the large density difference between the air and water, the air flow on top of the water column was neglected.

Naghashnejad and Shabgard/ Preprint submitted to peer reviewed international journal## 4. Results and Discussion

In order to validate the developed model, the numerical results were compared with those obtained form available analytical models for both liquid rise and meniscus profile. The analytical models are obtained by applying the following momentum balance to a control volume containing the liquid in the capillary channel:

$$\frac{\partial}{\partial t}\int_V \rho \vec{v}\, dV + \int_A \rho \vec{v}_b\, (\vec{v}_b \cdot \hat{n}) dA = \sum \vec{F} \qquad (5)$$

where the first and second terms on the left hand side show the rate of change of momentum of the liquid column and the net momentum crossing the boundaries of the control volume (liquid column), respectively, and the right hand side shows the sum of the forces acting on the control volume. In the surface integral, $\vec{v}_b$ is the fluid velocity at the boundary of the control volume relative to the boundary. Noting that the momentum only crosses the boundary at the column entry, Eq. (5) can be rewritten as:

$$\rho w \frac{d}{dt}\left[h(t)\frac{dh(t)}{dt}\right] - \rho w v_b^2 = 2\sigma\cos\theta - \rho g w h(t) - \frac{12\mu}{w}h(t)\frac{dh(t)}{dt} \qquad (6)$$

where $h$ is the rise of liquid column, $\sigma$ is the surface tension at the liquid-air interface, and $\theta$ is the contact angle. The right-hand side terms of the above equation are surface tension, gravitational, and viscous forces, respectively. The viscous force is obtained by assuming fully developed laminar flow between parallel plates. Determination of the relative velocity at the entrance of the capillary channel, $v_b$, requires further attention. As noted by Levine et al. [6], there are two possible scenarios to deal with the flow velocity at the inlet boundary. In the first scenario, it is supposed that the liquid in the reservoir is stagnant and speeds up with a fully developed velocity profile only *after* entering to the capillary. Thus, the second term on the left-hand side of the Eq. (6) vanishes.

The second scenario assumes that the liquid in the reservoir has zero velocity except the liquid element which enters the capillary at time $t$ with an average velocity of $dh/dt$. In this case, the right-hand side of Eq. (6) simplifies to:

$$\rho w h(t)\frac{d^2 h(t)}{dt^2} \qquad (7)$$

The above treatments of the inertial terms lead to two analytical solutions that are identical in gravity, viscous, and capillary forces and only differ in the inertia term. In the following, analytical solutions referred to as *"Analytical 1"* and *"Analytical 2"* correspond to the first and second aforementioned scenarios, respectively. It is noted that these scenarios are the limiting cases and the actual momentum entering the capillary conduit is expected to lie in between these two bounds. In the next section, the computational results for the capillary rise are compared with both the above-mentioned analytical solutions.

The independency of the computational results from the grid and time step size were verified by systematically varying the mesh resolution and time step size for the same input parameters. It was found that refinement of the grid beyond 50 cells in the transverse direction across the channel and decreasing the time step size less than 1e-5 s did not yield noticeable improvement of the results. Hence, the computational grid consisting of 50 cells in the transverse direction across the channel (~33,800 total cells) and a time step size of 1e-5 s were chosen for the numerical tests.

## 4.1 Rise of Water Column

Figure 2 shows the computational results for the rise of water column vs. time compared with two analytical solutions. As evident, the maximum rise of the water column predicted by the CFD model lied in between the two analytical solutions, and all three models predicted a similar qualitative trend. The capillary height showed an oscillatory behavior with a damping amplitude and eventually stabilized at an equilibrium height (Jurin height). An excellent agreement was observed between the computational and analytical results for the steady state capillary height. As indicated in Table 2, the equilibrium capillary pressure predicted by the CFD model shows a very good agreement with that obtained from the Young-Laplace equation, with an error of less than 1%.

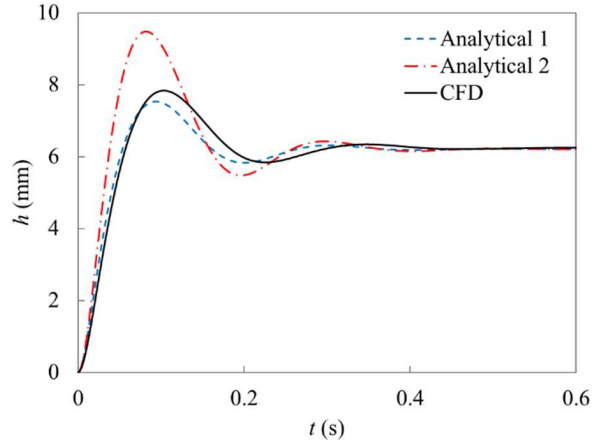

Fig. 2. Capillary rise of water column between parallel aluminum plates vs. time for wall spacing of $w = 0.7$ mm.

Table 2. Comparison between the simulation results and the Young-Laplace equation results for the equilibrium capillary pressure and the meniscus rise for wall spacing of $w = 0.7$ mm.

|           | Capillary pressure (Pa) | Height (mm) |
|-----------|-------------------------|-------------|
| Analytical | -60.813                | 6.210       |
| CFD model | -61.023                 | 6.233       |
| Error (%) | 0.35                    | 0.37        |

The time evolution of the capillary height for various wall spacing values is shown in Fig. 3. It can be seen that as the wall spacing increased, the equilibrium height decreased, and a longer time was needed to reach equilibrium. This trend is consistent with the previously reported data [15]. For the smallest wall spacing, a monotonic behavior was observed. Furthermore, it can be seen in Fig. 3 that by increasing the capillary spacing, oscillations occurred around the equilibrium height. These oscillations are mainly driven by the interplay between the inertial and gravitational effects, both of which are volumetric phenomena. On the other hand, the damping viscous forces are proportional to the surface area between the liquid and solid walls. Increasing the wall spacing results in a greater volume-to-surface ratio that implies a smaller damping effect compared to the forces that drive the oscillations.

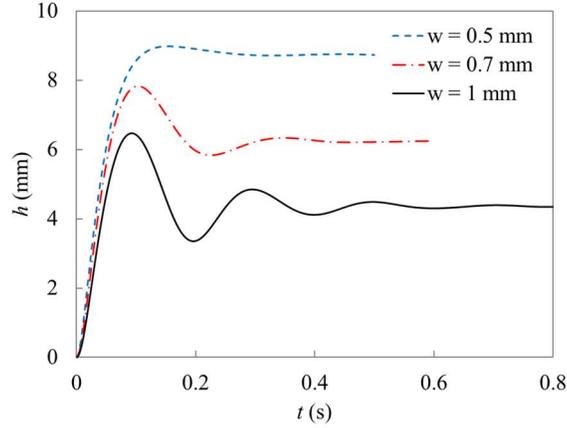

Fig. 3. Computational results for the capillary rise of water between vertical parallel aluminum plates vs. time for various wall spacings.

### 4.2 Meniscus Profile

The accuracy of the present computational model in predicting the steady-state meniscus profile was also verified by comparison with the analytical results from the model presented by Bullard et al [9]. Figure 4 shows the comparison results. The origin of the coordinate system in Fig. 4 was set to the contact line of the liquid and wall (left corner). As evident, the CFD results are in excellent agreement with the analytical solution, demonstrating the remarkable accuracy of the present model in predicting the meniscus profile.

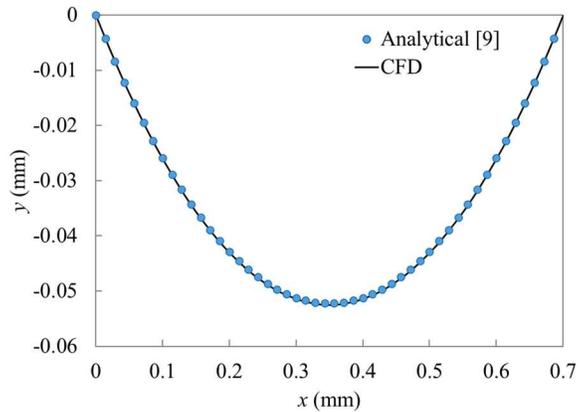

Fig. 4. Steady-state meniscus profile for water confined between parallel aluminum plates with $w = 0.7$ mm.

### 5. Conclusions

A computational model was developed to simulate the capillary rise of liquid between vertical parallel plates. A dynamic mesh method was employed to directly track the interface with no need for implicit interface tracking schemes. An excellent agreement was observed between the numerical results and the analytical solution for the steady-state capillary height and the meniscus profile. Also, the results for various wall spacing values demonstrated different flow regimes from monotonic to oscillatory. An important feature of the present CFD model is the sharp interface tracking and reconstruction. This feature is particularly advantageous in problems involving interfacial heat and/or mass transfer where accurate specification of the interfacial fluxes is critical.